\documentclass[reprint,amsmath,amssymb,aps,showpacs]{revtex4-1}
\usepackage{graphicx,color}
\usepackage{dcolumn}
\usepackage{bm}

\newcommand{\figref}[1]{Figure~\ref{#1}}
\newcommand{\secref}[1]{Section~\ref{#1}}

\newcommand\ion[2]{#1$\;${\small\rmfamily\@Roman{#2}}\relax}%
\newcommand{\C}[1]{\ensuremath{{}^{#1}{\rm C}}}
\newcommand{\0}[1]{\ensuremath{{}^{#1}{\rm O}}}
\newcommand{\Ne}[1]{\ensuremath{{}^{#1}{\rm Ne}}}
\newcommand{\Ni}[1]{\ensuremath{{}^{#1}{\rm Ni}}}
\newcommand{\Co}[1]{\ensuremath{{}^{#1}{\rm Co}}}
\newcommand{\Fe}[1]{\ensuremath{{}^{#1}{\rm Fe}}}

\newcommand{\pv}{\ensuremath{\phi}}

\newcommand{\unitspace}{\ensuremath{\,}}
\newcommand{\usp}{\unitspace}

\newcommand{\unitstyle}[1]{\ensuremath{\mathrm{#1}}}
\newcommand{\power}[2]{\ensuremath{{#1}^{#2}}}

\newcommand{\centi}{\unitstyle{c}}
\newcommand{\kilo}{\unitstyle{k}}

\newcommand{\meter}{\unitstyle{m}}

\newcommand{\second}{\unitstyle{s}}

\newcommand{\cm}{\centi\meter}
\newcommand{\gram}{\unitstyle{g}}

\newcommand{\grampercc}{\gram\usp\power{\cm}{-3}} 

\newcommand{\kms}{\kilo\meter\unitspace\power{\second}{-1}}

\newcommand{\Msun}{\ensuremath{M_\odot}}

\newcommand{\rhoDDT}{\ensuremath{\rho_{\rm DDT}}}
\newcommand{\COreac}{\ensuremath{\C{12}\left(\alpha,\gamma\right)\0{16}}}

\begin{document}

\title{The influence of chemical composition on models of Type Ia supernovae}

\author{Alan C.\ Calder}
\affiliation{
Department of Physics \& Astronomy \\
The State University of New York - Stony Brook, Stony Brook, NY, USA}
\affiliation{
New York Center for Computational Sciences \\
The State University of New York - Stony Brook, Stony Brook, NY, USA}
\author{Brendan K.\ Krueger}
\altaffiliation[Presently at ]{CEA Saclay, France}
\affiliation{
Department of Physics \& Astronomy \\
The State University of New York - Stony Brook, Stony Brook, NY, USA}
\author{Aaron P.\ Jackson}
\affiliation{
National Research Council Research Associateship Program}
\affiliation{
Laboratories for Computational Physics and Fluid Dynamics, 
Naval Research Laboratory, Washington, DC, USA}
  
\author{Dean M.\ Townsley}
\affiliation{
Department of Physics and Astronomy \\
The University of Alabama, Tuscaloosa, AL, USA}

\begin{abstract}
Type Ia supernovae are bright stellar explosions distinguished by 
standardizable light curves that allow for their use as distance 
indicators for cosmological studies. Despite the highly successful 
use of these events in this capacity, many fundamental questions 
remain. Contemporary research investigates how 
properties of the 
progenitor system that follow from the host galaxy such as composition and age
influence the brightness of an event with the goal of better understanding 
and assessing the intrinsic scatter in the brightness.
We provide an overview of these supernovae and proposed progenitor systems,
all of which involve one or more compact stars known as white dwarfs. 
We describe contemporary research 
investigating how the composition and structure of the progenitor 
white dwarf systematically influences the explosion outcome assuming
the progenitor is a single white dwarf that has gained mass from a companion.
We present results illustrating some of these systematic effects from our research.
\end{abstract}



\maketitle

\section{Introduction}
\label{sec:intro}

\subsection{Historical Overview of Supernovae}

``New'' or ``guest'' stars have occasionally appeared in the sky 
since time immemorial. Ancient observations include solar system objects 
such as comets  and stellar explosions that appear suddenly
as a newly visible star. The earliest record of an event associated
with a stellar explosion is found the {\em The Book of the Later Han}
which describes the appearance of a ``guest star''  in the year
185~\citep{stothers77,supernova88}. Modern study of stellar explosions
began with Tycho Brahe's naked-eye observations of the event we now
call Supernova 1572, and the name ``nova'' applied to the phenomena
of stellar explosions follows from his book {\em De Stella Nova} 
(``On the New Star''). Tycho's observations 
were of such fidelity that later astronomers were able to reconstruct the 
light curve and color evolution of this event for comparison and 
classification~\citep{badde45,rl2004}. 

By the early twentieth century, technological
advances in astronomy led to recognition of the vast distances to
other galaxies, implying that novae observed in other galaxies were
in fact extremely bright. This advance, together with advances in
theory, resulted in the classification of ``super-novae'' as bright explosions
associated with the violent death of a star~(Ref.\ \citealp{bz34}; see also Refs.\
\citealp{hillebrandt.niemeyer:type} and \citealp{osterbrock2001} for discussion
of pioneering work).
Observations of a supernova typically produce spectra and a light curve, with
spectra taken during the course of an event providing information identifying 
the constituent elements and the light curve plotting the intensity of 
light vs.\ time.  
As originally proposed by~\citet{minkowski41}, 
supernovae are classified by properties of their spectra and light 
curves, with the principal distinction arising between events 
without evidence of hydrogen in their spectra (Type I) and events with evidence
of hydrogen in their spectra (Type II). Type I supernovae were further
divided into Types Ia, Ib, and eventually Ic~\citep{bertola64,porterfilippenko87,
wheelerharkness1990conf,Fili97}.  The Type Ia sub-classification depends on the 
observation of a specific Si line~\citep{Fili97,hillebrandt.niemeyer:type} 
not seen in Types Ib and Ic. Type Ib and Ic are distinguished by Type Ib spectra exhibiting 
He features while Type Ic spectra do not.

Given the extreme brightness of supernovae, only two possible
energy sources are thought to account for these phenomena: the release
of gravitational potential energy or the release
of nuclear binding energy. Events in all of the observational classifications, 
with the exception of Type Ia, are understood to follow from the gravitational collapses 
of a massive star that has exhausted its nuclear fuel~\citep{colgatewhite66}. The remnant 
in this type of explosion is a neutron star or black hole, of which many have been observed.
Compact remnants have never been observed from Type Ia supernovae. While the classification 
follows from the spectral signature, Type Ia supernovae are now understood to be the result 
of a thermonuclear explosion consuming roughly one and a half solar masses of degenerate 
stellar material composed principally of C and O~\citep{hillebrandt.niemeyer:type}. The 
progenitor systems of these events, however, remain the subject of debate. Most proposed 
Type Ia progenitor systems follow from the suggestion of 
\citet{hoylefowler60} of a thermonuclear runaway occurring in the
core of a star supported by electron degeneracy. Various settings in 
which suitable conditions for the explosive burning of degenerate
stellar material have been explored, and scenarios typically
involve one or more white dwarf stars, as we discuss below.

\subsection{Observations of Type Ia Supernovae}

The release of nuclear binding energy via explosive burning of degenerate
C and O provides more than enough kinetic energy to unbind a white dwarf
star, and the expansion velocities of freshly synthesized elements 
typically reach on the order of $10,000~\kms$.
The peak brightness of a Type Ia supernova is set 
not by the explosion energy, but
by the synthesis in the explosion of radioactive \Ni{56}, which decays
via the chain \Ni{56} to \Co{56} to \Fe{56} releasing photons that power
the observed light curve \citep{pankey62,colgatemckee69,colgatepetschekkriese80,
kuchneretal94}.

The vast majority of Type Ia supernovae obey a correlation in which the 
peak brightness is positively correlated with the
timescale over which the lightcurve decays from its maximum~\citep{pskovskii77}.
This ``brighter is broader'' trend is known as the Phillips
relation~\citep{phillips:absolute} and it allows the
peak brightnesses to be calibrated so that these events may be treated
as ``standard candles" for determining distances.
The resulting parameter describing the scaling is known as
$\Delta m_{15}(B)$, which is the change in the apparent $B$-band
magnitude from peak brightness to 15 days later.
The correlation is understood physically as stemming from having both the 
luminosity and opacity being set by the mass of \Ni{56} synthesized in the
explosion~\citep{arnett:type,pinto.eastman:physics,Kasen2007On-the-Origin-o}.
This relation has been exploited to make Type Ia supernovae the premier
distance indicators for cosmological studies. We note that because the
galaxy/stellar population is known to have evolved significantly since
high ($z \approx 1$) redshifts, the systematics of this evolution must
be accounted for to allow precision cosmological measurements. Avoiding
relying on empirical relationships is one important goal of the 
theoretical study of Type Ia supernova explosions.

While the yield of \Ni{56} explains the first-order variations in the light
curve, current research also aims to understand higher-order effects and
the physics behind the Phillips relation.
Observations report that Type Ia supernovae appear to have an intrinsic scatter 
of a few tenths of a magnitude after calibration, forcing a minimum uncertainty 
in any distances measured by using Type Ia supernovae as standardizable
candles~\citep{JacobyEtAl92, Kirshner09}.  An important goal of theoretical
research into Type Ia supernovae, from the standpoint of cosmology, is to understand 
the sources of scatter and to identify potential systematic biases by studying the
effects of various properties on the mechanism and nucleosynthetic yield of the
supernova. The surrounding stellar population and various properties of the progenitor
such as its composition, zero-age main sequence mass, thermodynamic state 
prior to ignition, and cooling and accretion history are known to affect
the lightcurves of Type Ia supernovae; the role of these ``secondary'' parameters is the
subject of considerable study
(e.g., Refs.\ \citep{Roepetal06_2, hoeetal2010,seitenzahletal11}).
Additionally, many of these effects may be interconnected in complex
ways~\citep{DomiHoefStra01, Lesaffre2006The-C-flash-and, townetal2009,mengyang2011}.

\subsection{Trends in Observations of Type Ia Supernovae}

The interest in Type Ia supernovae that follows from the 
successful use of these events as distance indicators in studies revealing 
the acceleration of the 
Universe~\citep{riess.filippenko.ea:observational,perlmutter.aldering.ea:measurements} 
has led to modern observational campaigns designed to explore systematic 
effects on the brightness of these events~\citep{Sullivan2006Rates-and-prope,sullivanetal2010,
lietal2011c}. 
Observations explore correlations 
between observed properties of an event such as peak brightness and properties of 
the host galaxy such as its composition, age,
and mass. Many of these galactic properties are correlated, which makes the
task of determining the underlying physical reason for correlations between
properties of supernovae and their host galaxies difficult. Compounding this
difficulty is the fact that
detailed observations of galaxies are not available at high redshifts, and
parameters (typically galaxy mass and star formation rate) are therefore
inferred from galactic models that reproduce the
observed spectral energy distribution.

\citet{Hamuy1996} reported that the peak brightness
correlates with the morphological type of the host galaxy, with the most
luminous supernovae tending to be hosted by late-type, spiral galaxies.
While the trend is striking, supernovae within the same morphological
type do not necessarily share the same physical environment. Several
physical properties of galaxies tend to correlate with the morphological
type. Early-type, elliptical galaxies are thought to form through
galaxy mergers and are more massive, contain older stellar populations,
and have little star formation. 
The composition of a galaxy depends largely on the proportion of 
material that has previously been processed in
stars (i.e.\ elements other than H and He, which are
collectively referred to as ``metals''). ``Metallicity,'' the
relative abundance of these elements, in the gas phase is correlated 
with the galaxy mass as galaxies with deeper gravitational potential wells
tend to more effectively retain metals~\citep{Tremonti2004}. Therefore,
early-type galaxies also tend to have higher metallicities. Late-type,
spiral galaxies are less massive, are actively forming stars, and hence
contain younger stellar populations. Likely some combination
of these properties bias the progenitors of Type Ia supernovae to produce
the observed correlation with galaxy morphology.

Despite challenges, \citet{GallGarnetal05}
observed a slight dependence of peak brightness on metallicity
with dimmer events in metal-rich galaxies. A conclusive
trend, however, has proven elusive. Due to the weak correlation with metallicity,
progenitor age is suspected to be primarily responsible for variations
in the peak brightness~\citep{GallGarnetal05,gallagheretal+08,neilletal+09,
howelletal+09}; however, challenges remain in drawing this conclusion.
First, the present gas-phase metallicity of the host galaxy may not
be representative of the metallicity of the gas the progenitor star formed from,
particularly for older stellar populations. Second, the determination of the
age of a stellar population from the integrated light of a galaxy
is ``degenerate'' with the determination of the metallicity~\citep{silvaelston94,
worthey94}. That is, both effects tend to redden the light from a galaxy, 
so it is difficult to distinguish the true cause of the reddening.
For observations of high-redshift galaxies, the metallicity must be inferred
from the galaxy mass, precluding the possibility of directly determining its
effect on the supernova light curve~\citep{howelletal+09}.
When correlations with the inferred stellar population age are considered,
dimmer supernovae are found in older populations. 

Metallicity could also be
responsible for variations in the peak brightness of {\it calibrated\/} light curves.
At present, observers do not use the Phillips relation \citep{phillips:absolute} but
employ much more sophisticated methods to calibrate light curves and thereby measure
extragalactic distances. These methods construct calibrated light curves in multiple pass-bands
based on templates from nearby events and include corrections for extinction, time dilation, and the
K correction for shifting of spectra due to cosmological redshift.  Examples include the Spectral Adaptive 
Light Curve Template SALT \citep{guyetal2005,guyetal2007}, the Multicolor
Light-Curve Shapes method MLCS2k2, \citep{jha2002,jhareisskirshner2007}, and the (Python-based) supernova
light curve fitter SNooPy \citep{burnsetal2011}.

Any metallicity dependence reported for calibrated light curves
is representative of systematic effects that have not been taken into account in the calibration
procedure. \citet{GallGarnetal05,gallagheretal+08} report a dependence in the peak
brightness in the calibrated light curves with metallicity, suggesting
the possibility of systematic effects with redshift that would have implications
on cosmological parameters and the equation of state of Dark Energy. \citet{howelletal+09}, 
however, using a different calibration method, find no such dependence
on metallicity.

Theoretically, the presence of metals in
a progenitor white dwarf influences the outcome of the explosion by
changing the path of nuclear burning, which influences the amount of
$^{56}$Ni synthesized in an event~\citep{timmes.brown.ea:variations}.
Because the decay of $^{56}$Ni powers the light curve, metallicity
can therefore directly influence the brightness of an event.
Metallicity can also influence the outcome of an explosion in other ways,
including changes in the structure of the white dwarf that result from
the influence of metallicity on stellar evolution, sedimentation within the
white dwarf, the nuclear flame speed, additional sources of
opacity, and properties of the thermonuclear burning such as the ignition density (described
below)~\citep{Nomo84,DomiHoefStra01}. 

In addition to properties of the light curves, Type Ia supernovae rates
may also yield clues to the progenitors of these explosions. Unlike
other types of supernovae, Type Ia have a non-zero rate in older populations,
which also points to a white dwarf progenitor. Some results indicate
that the dependence of the Ia rate on stellar age
is best fit by a bimodal distribution having a prompt component 
less than 1~Gyr after star formation and a tardy
component several Gyr later~\citep{MannucciEtAl06,RaskinEtAl09}, with the
prompt component appearing brighter on average than the tardy component.
A bimodal distribution with delay time or mean stellar age would imply
two distinct progenitor formation channels.
Other studies only indicate a correlation between the delay time and the
brightness of Type Ia supernovae favoring a single formation channel,
with dimmer events occurring at longer delay
times~\citep{gallagheretal+08,howelletal+09,neilletal+09,BrandtEtAl10}.

Perhaps the most challenging aspect of observing supernovae is accounting for
dust. The process of calibrating light curves and inferring luminosity 
distances relies on using the width-luminosity relation and color corrections, and the manner in which
color is linked to dust is considered controversial~\cite{chotardetal2011}.
Color variations observed between supernovae correlate with the
dust content of the host galaxy, but it is also possible that some of the
color variation is intrinsic to properties of the explosion. Observationally
disentangling intrinsic variability from effects of dust is the subject of active 
research~\cite{chotardetal2011}.

Observations find a surprising range in the variation of the 
intrinsic brightness of these events, including very bright events that 
suggest more material burned than could originate from a single white 
dwarf~\citep{howell+06,scalzo+10,yuan+10,tanaka+10}. The logic of observations
suggesting that more material burns than could come from a single white
dwarf is based on the fact that a white dwarf is supported by electron degeneracy
and is therefore subject to a maximum mass, known as the Chandrasekhar limit,
beyond which it will collapse to a neutron star~\citep{chandra31,chandra35,chandra67}. 
A bright observation that suggests more \Ni{56} synthesized than the Chandrasekhar 
maximum mass therefore implies more than one white dwarf was involved. Accordingly, 
contemporary research explores the efficacy of several progenitor systems with one 
or more white dwarfs.

\subsection{Proposed Progenitors Systems of Type Ia Supernovae}

We briefly introduce some proposed progenitor systems of Type Ia supernovae 
and outline contemporary research in each. The accepted result for a Type Ia
supernova is an event that produces about $0.6 \Msun$ of \Ni{56}, the decay
of which produces the light curve, and most models posit the thermonuclear burning
of a mix of C and O under degenerate conditions 
(see Refs.\ \citep{mortsellbergstrom2002,biermannclavelli2011}
for examples of alternative models.)
Reviews of progenitor models may be found in 
\citet{hillebrandt.niemeyer:type,livio2000,wanghan2012}. While we note that events 
produce $0.6 \Msun$ of \Ni{56}, there is of course scatter in the \Ni{56} yield 
(and brightness) of events and sub-classifications, e.g. Branch-normal, 1998bg-, 1991T-, 
and 2002cx-like events~\citep{branchfishernugent93,stritzinger2006,lietal2011b,foleyetal2013}.

The association of Type Ia supernovae with white dwarfs began
with the work of \citet{hoylefowler60} who found that
the thermonuclear incineration of a mixture of C and O under
degenerate conditions could explain Type I supernovae. \citet{hoylefowler60}
theorized a mass approaching the Chandrasekhar limit would be required, but the question
remained as to where an appropriate massive stellar core might be found.
A suitable progenitor system must satisfy two requirements. First, the system
must have enough mass to produce the amount of intermediate-mass 
and Fe-group elements, e.g.\ \Ni{56}, inferred from observations, a small
fraction of which
imply more synthesized mass than could come from a single white dwarf. Second,
while massive white dwarfs may be composed of mixtures of C and O or
O, Ne, and Mg, the progenitors of Type Ia supernovae  must be composed principally of C and O because
the temperatures required to ignite O or Ne cannot be achieved prior to
collapse (see Ref.\ \citep{mueller91} and references therein). Accordingly,
proposed progenitor systems typically involve either multiple C-O white dwarfs or mass
transferred from a companion onto a C-O white dwarf.

Many questions concerning progenitor channels remain and are actively being
explored as described below. Constraints on progenitor channels are also under study from
the perspective of population
synthesis (see for example Refs.\ \citep{yungelsonlivio2000,wanglihan2010,
ruiterbelczynskifryer2009,ruiteretal2011}).
We note that many questions remain concerning the fundamentals of thermonuclear
burning in white dwarf material \citep{bravoetal2011}, and, as mentioned
above, disentangling the primary from secondary parameters in
the observed light curve is a challenge. These many
issues influence the question of Type Ia progenitor channels.

\subsubsection{A Single Massive White Dwarf}
\label{sec:single}


This progenitor channel is referred to as a ``single degenerate" due to the basic 
requirement that the white dwarf accretes material from a non-degenerate companion 
such as a red giant, helium, or main sequence star \citep{whelaniben73} (see
also Refs.\  \citep{hachisukatonomoto1996,hanpodsiadlowski2004,wanghan2010a,wangetalmnras2009}). 
In fact, the single degenerate 
channel may be comprised of more than one evolutionary pathway, and  several are actively 
being investigated. While no particular channel has direct observational evidence 
supporting it, some evolutionary pathways are scrutinized more heavily due to the 
lack of expected observational features in the light curve and spectra. In particular, 
a red giant companion to an exploding white dwarf is expected to produce a detectable 
feature in the UV when the blast wave interacts with the red giant gas 
\citep{kasenetal2010}. To date, observations of this type of interaction 
are lacking, which rules out the possibility that the red giant channel is the only 
evolutionary path. It may still be possible, though, that this pathway 
contributes a small percentage of the total rate of Type Ia supernovae. 

Some specific observations of Branch-normal 
events provide upper limits of the mass of the companion of those particular systems. 
For example, the recent SN 2011fe is the closest Type Ia supernova to explode in modern times 
allowing the earliest time observations as well as detailed pre-explosion images that 
provide an upper limit to the brightness of the companion star
\citep{NugentEtAl2011,brownetal2012}. These observations have 
ruled out luminous red giants and almost all helium stars as the companion to this 
particular system \citep{LiEtAl2011}. \citet{Wheeler2012} 
has suggested M-dwarfs as a viable donor star due to the expected rates at which 
WD/M-dwarf binaries are produced; however, further research is needed to explore 
the unique accretion process that is assisted by the magnetic properties of both the 
M-dwarf and white dwarf. The evidence in PTF 11kx \citep{dildayetal2012} of nova 
shells predating the supernova explosion implies that at least some fraction of 
fairly normal Type Ia supernovae must originate from single degenerate type systems.

The single degenerate W7 model
by \citet{Nomo84} that used a parameterized burning velocity
offered good agreement with observations and strongly motivated further study
of the single degenerate progenitor. The key to the success of W7 follows
from the fact that the density at which degenerate stellar material burns largely
determines the composition of the ash. At high densities ($\rho \gtrsim
10^7 \grampercc$), nuclear fuel reacts by fusion all the way to
the Fe group on the timescale of the burning front propagating
though the white dwarf. At lower densities, the burning is incomplete and stops
at the Si group or even after C burning. The parameterized
reaction of W7 allowed the star to expand prior to the completion of
burning, which produced a stratified remnant that agreed well
with observations (see Fig.\ 4 of Ref.\ \citep{Mazzetal08}).

As the success of the W7 model demonstrates, the nature of the burning in the 
single degenerate progenitor scenario is critical to the explosion outcome. \citet{arnett69} 
investigated C detonations, a case in which burning proceeds as a supersonic front, 
in the C-O cores of massive stars.  The result was that the entire core was 
incinerated to the Fe group, which is now understood as producing too much 
Fe. The alternative to a detonation is a deflagration, a subsonic burning 
front that, in the case of white dwarf material, proceeds by the conductive
propagation of heat. Factors such as the interaction of the flame with turbulence
influence the burning rate, but after considerable study, the consensus is 
that a pure deflagration will not produce a normal Type Ia 
supernova because the 
relatively slow burning of the deflagration expands the star too much to
match observed abundances (see Ref.\ \citep{roepkeetal07} and references therein).

The favored single degenerate explosion model is that of a period
of deflagration (subsonic burning) that allows the star to expand
followed by a detonation (supersonic burning) that rapidly 
incinerates the expanded star. This delayed detonation paradigm
was introduced by \citet{khokhlov91+dd}, and simulations of this class 
of models, like Nomoto's W7, demonstrate good agreement with 
observations \citep{hoflich.khokhlov.ea:delayed,HoefKhok96} and have
been widely accepted. Many variations on this theme exist, however, including 
pulsational detonations in which the white dwarf expands and
recollapses, subsequently igniting a 
detonation~\citep{arnettlivne94a,arnettlivne94b,HoefKhok96}, gravitationally 
confined detonation (GCD) in which compressional heating ignites a detonation 
in material flowing across the surface of the white 
dwarf \citep{PlewCaldLamb04,Jordan2007Three-Dimension} (see also Refs.\ \citep{Seitetal09det,chamulaketal2012}),
and models in which a deflagration transitions to a detonation when
the right local conditions are met \citep{1986SvAL, woosley90, khokhlov91+dd, 
hoflich.khokhlov.ea:delayed, HoefKhok96, KhokOranWhee97, NiemWoos97, hwt98, Niem99}.

Questions remain about the details of flame ignition: does a
localized region initiate explosive burning or is it distributed?
If localized regions runaway, how many are there and how are they distributed?
Due to the highly nonlinear evolution of the deflagration phase,
the details of the ignition process are critically important for
determining the outcome of an explosion. At one extreme, single, off-center
ignitions likely lead to an asymmetric deflagration phase, like that described
in the GCD scenario, while distributed central ignitions lead to fairly
symmetric deflagrations. A recent study by \citet{Maedaetal2010Nature}
suggested that distributed, off-center ignitions produce viewing
angle effects that influence the peak brightness, and may be largely
responsible for the observed variation.

\subsubsection{Merging White Dwarfs}

The idea of merging white dwarfs as the progenitors of Type Ia supernovae 
has been around for some time with investigations into near- and super-Chandrasekhar
mass coalesced objects \citep{tutukovyungelson76,tutukovyungelson79,webbink84,ibentutukov84}.  The simplest picture of 
the merging white dwarf scenario has tidal effects disrupting the 
secondary, which forms a disk from which the primary subsequently 
accretes C and O.  One of the most commonly cited perils of the
merging white dwarf model is that the C will ignite at the edge of
the coalesced object, and a flame will burn inward, converting the
C-O white dwarf into an O-Ne-Mg white
dwarf~\citep{saionomoto1985,saionomoto2004}.  Instead of becoming a
Type Ia supernova, further accretion leads to the collapse of the white dwarf
into a neutron star, a scenario termed ``accretion-induced collapse''
\citep{nomotokondo1991}.
Because of this concern, the review of \citet{hillebrandt.niemeyer:type} 
dismissed the double degenerate scenario.  Such a collapse
can be avoided if the accretion rate is low enough ($<$ few
$\times 10^{-6}~M_\odot~\mathrm{yr}^{-1}$~\citep{kawai1987,saionomoto2004})
that the heating at the surface of the primary does not ignite
the C.  Models of violent mergers of white dwarfs (described below) 
also can avoid an accretion-induced collapse by initiating a 
detonation of the sub-Chandrasekhar mass primary~\citep{pakmoretal2011}.
Accordingly, there is great interest in demonstrating that
the detailed accretion and burning processes provide a mechanism 
to avoid accretion-induced collapse. 

There are a number of studies reported in the literature
that have advanced the state of understanding of white dwarf
mergers that we highlight here.
Other studies investigated collisions of 
white dwarfs~\citep{rosswogetal2009,raskinetal2009,raskinetal2010,lorenaguilaretal2010}.
These events may produce Type Ia supernovae, but because the  
likely locations for such
collisions are in the dense cores of globular clusters, these
events will be rare, perhaps as few as 10--100 per year in the
local ($z < 1$) universe \citep{raskinetal2009}.
There is also a core-degenerate merger scenario in which a white dwarf 
merges with the core of an asymptotic giant branch 
star (see Ref.\ \citep{mengyang2012} and references therein).

\citet{benzetal90} performed some of the earliest simulations 
of merging white dwarfs and found that the merger remnant is best 
described as a hot, rapidly rotating thick disk consisting of the 
tidally-disrupted secondary surrounding the original primary white dwarf.  
These simulations confirmed the basic idea that the inspiral of the white
dwarfs driven by gravitational radiation can cause the ultimate merger
of the system, and emphasized that angular momentum transport is the
primary physical mechanism that shapes the outcome.

\citet{yoonetal2007} modeled the merger of $0.9~M_\odot$ and
$0.6~M_\odot$ white dwarfs and argued that accurate modeling of the 
merger event is critical to determining the details of the coalesced 
remnant.  Their simulation indicated the less
massive star was disrupted, and the result was  described
as ``a differentially rotating single C-O star consisting of a slowly
rotating cold core and a rapidly rotating hot extended envelope
surrounded by a Keplerian disk'' instead of the white dwarf and thick 
disk found in earlier studies. Based on mapping the result of the 
three-dimensional simulation to one dimension to study burning,
they argued that this configuration does not give rise to ignition at the edge of the
star and hence the formation of an O-Ne-Mg white dwarf. Instead, central
ignition of the C-O remnant can occur.

An extensive study was made by~\citet{lorenaguilaretal2009},
who considered different systems with varying masses.  As with 
the Yoon et al.\ case, they found that the systems avoid
igniting the C explosively at the instant of merger. They 
found a coalesced remnant configuration similar to that of Yoon et al.,
again consisting of a rotating core, hot envelope (which they term
``corona''), and a Keplerian disk.  Perhaps owing to
different initial conditions, the final core temperatures differed by orders
of magnitude between the two groups. \citet{lorenaguilaretal2009} concluded that 
the disks surrounding the remnant are very turbulent, and argued that
this would lead to high accretion rates and, subsequently, 
off-center ignition and accretion induced collapse.  
\citet{Shenetal12} also find that C does 
not explosively ignite during the merger, but for different physical reasons owing 
to the ability of magnetic stresses to redistribute angular momentum. Over a longer 
timescale after the merger, they find that conditions at the interface of the 
tidally disrupted secondary and primary white dwarf are such that convective 
C burning should ensue leading to slow shell burning that eventually results in a
high-mass O-Ne-Mg white dwarf or accretion induced collapse.

Other work serves to stress the importance of the fidelity of 
numerical methods and included physics. \citet{fryerdiehl2008} 
and \citet{diehletal2008} performed simulations of merging white
dwarfs with similar methods but found different results.  The collaboration 
of \citet{dsouza2006} and \cite{motl2007} with different, grid-based
methods found that, depending on the initial configuration, stable
mass transfer could be achieved, a result seen only 
by \citet{fryerdiehl2008}. Recently, \citet{danetal2011}
investigated the issue of stable mass transfer and found it
possible with a careful treatment of initial conditions and \citet{danetal2012}
performed a wide-ranging survey, but found the chance of a successful
detonation following a merger of C-O white dwarfs small. A nice opus 
on many of the concerns was produced by \citet{pakmoretal2012b}.

Recent studies indicate success with white dwarf merger models.
\citet{pakmoretal2010} found that the case of nearly equal masses 
of $\approx 0.9 \Msun$ may explain sub-luminous events. \citet{pakmoretal2011} found that 
violent mergers with primary masses of $\approx 0.9 \Msun$  and mass ratios
$< 0.8$ will produce explosions and are promising candidates for sub-luminous events.
Very recently, \citet{pakmoretal2012a} found that under the right conditions,
the violent merger of two C-O white dwarfs may produce a normal 
Type Ia supernova.

The merger of two white dwarfs is also thought to possibly produce
a rapidly rotating massive white dwarf \citep{steinmetzmullerhillebrandt92}.
Following considerable exploration of detonation scenarios including
one-dimensional rotating and off-center models (see Ref.\ \cite{mueller91} and references therein), the first multidimensional
simulations of detonations in rapidly rotating white dwarfs were performed by 
\citet{steinmetzmullerhillebrandt92}. The result of these two-dimensional
simulations indicated that as with earlier studies of detonations \cite{arnett69},
the bulk of the star burned all the way to Fe-group elements and did
not produce the expected intermediate-mass elements. 

Recently \citet{pfannes2010a} performed full three-dimensional 
simulations of turbulent deflagrations in rapidly rotating massive
white dwarfs. They found that the amount of burning in their 
rotating models was similar to that of deflagrations occurring in
non-rotating massive white dwarfs, producing ``comparably weak
and anisotropic explosions that leave behind unburnt
material at the center and in the equatorial plane.'' The conclusion
drawn from their deflagration models is that rotation will not
explain the observed variation in peak luminosities.
\citet{pfannes2010b} explored detonations occurring in rapidly 
rotating massive white dwarfs with three-dimensional simulations.
In this case, models indicated that prompt detonations could produce 
bright supernova-like events, including producing sufficient amounts 
of intermediate-mass elements. They conclude that detonations in 
rapidly rotating massive white dwarfs can explain some bright 
events.

Additional support for the idea that rapidly rotating massive
white dwarfs might explain some bright events comes from
the recent study of \citet{hachisuetal2012}. They presented a model
that includes optically thick winds from an accreting white dwarf,
mass stripping from the binary companion by the white dwarf wind,
and support from differential rotation. They report that masses
can reach up to $2.7 \Msun$ and explored the results of
explosions occurring in three mass ranges that
depend on the character of the rotation and the onset of secular
instability and conclude that the model can explain both bright 
and normal events, and that differences in angular momentum loss
rates (and thus spindown rates) might explain ``prompt'' and
``tardy'' components of supernova observations.


\subsubsection{Double Detonation Models}

Another progenitor channel under study is known as the ``double detonation" model.
This idea is similar to the single degenerate scenario described in \secref{sec:single}, but 
in this case the white dwarf does not need to approach the Chandrasekhar 
limiting mass. Pioneering work introducing double detonation models was performed
by \citet{woosleyweavertaam80,taam80a,taam80b}
and \citet{nomoto80,nomoto82b}.
Studies indicated that the under the right conditions, an accreted He
layer could flash with such strength as to produce an inwardly propagating
shock that would ignite a detonation in the C-O core. Studies also indicated
that the double detonation scenario could work for a wide range of white dwarf 
masses, not just the near-Chandrasekhar case \citep{livne90}, and, accordingly, lower mass models were 
dubbed ``sub-Chandrasekhar'' \citep{ww94}.

Multidimensional simulations confirmed many of these findings, but as with other
models, uncertainties in parts of the problem like the initial conditions
were found to play a significant role on the explosion outcome. Accordingly, a 
common conclusion was that double detonation models may explain some 
events~\citep{livneglasner91, livnearnett95,HoefKhok96, hoeflichetal96, wigginsfalle97,
wigginsetal98,garciasenzbravowoosley99}. Most models included a massive He layer
$\sim 0.2 \Msun$ so that intermediate and heavy elements synthesized in the He detonation 
of such models would appear in the outer parts of the ejecta, a result at odds with 
observations~\citep{HoefKhok96, hoeflichetal96,finkhillebrandtroepke2007,simetal2010}.

Recently, however, \citet{bildstenetal2007} found that fairly thin He layers, 
$\sim 0.06 \Msun$, could become dynamically unstable and flash on sub-Chandrasekhar 
mass white dwarfs, and \citet{finketal2010} found that a He detonation at the base 
of the accreted layer ``robustly triggers'' secondary detonations in the C-O core. 
\citet{kromeretal2010} further investigated the models of \citet{finketal2010} and 
calculated light curves. They found the right range of rise and decline times and 
brightness, but found the color too red due to composition of the He shell ash.

\citet{woosleykasen2011} performed an extensive one-dimensional parameter study, 
calculating nucleosynthesis and light curves for a variety of white dwarf 
masses and temperatures and He accretion rates. They found that only hot, 
massive white dwarfs with thin He layers produce Ia-like events and discussed
a number of potential objections to these models as a general solution to
the Type Ia supernova problem.  
\citet{simetal2010} investigated central detonations 
in bare (no He layer) sub-Chandrasekhar mass C-O white dwarfs. They produced light curves
and detailed nucleosynthesis results and concluded that such models are ``in
qualitatively good agreement with observed properties of Type Ia supernovae,''
but note that the question of realistic progenitor systems in keeping with
these models remains. \citet{simetal2012} investigated double detonation scenarios
with both edge-lit and central C detonations. They found that for relatively
higher mass accreted He shells, conditions are likely for a core detonation to occur 
following a He detonation for all expected C-O core masses. The light curves of these 
core detonation models were readily distinguishable from edge-lit or He detonation
with no core detonation models. All these recent studies serve to stress the importance
of details of the accreted layer to the distribution of nuclei in the ejecta and
the need for high-fidelity spectral observations for validating this class
of models.


\section{Single Degenerate Delayed Detonation Models}

The single degenerate scenario posits compressional heating of 
a white dwarf due to accretion. The core of the white dwarf begins 
runaway heating due to C fusion outpacing neutrino losses 
\citep{nomoto82a,WoosWeav86} and a period of convection ensues 
\citep[][and references therein]{nonakaetal2012}. When heat generation 
outpaces transport of heat by convection, a flame is born that
eventually will incinerate the white dwarf. The central density at flame ignition 
is slightly decreased (see e.g.\ \citealt{PiroBild08}) and  is generally near
$\sim 2\times 10^9$~g~cm$^{-3}$ for successful models of Type Ia supernovae
(see e.g.\ Refs.\ \citep{Nomo84,HoefKhok96}).

A variation of the delayed detonation mechanism (within the single degenerate
paradigm) is that of a deflagration to detonation transition (DDT).
After ignition, the flame propagates as a subsonic deflagration for a while
and then transitions to a supersonic detonation that rapidly consumes
the star~\citep{1986SvAL, woosley90, khokhlov91+dd, hoflich.khokhlov.ea:delayed,
HoefKhok96, KhokOranWhee97, NiemWoos97, hwt98, Niem99}. We describe the
details of our implementation of this explosion mechanism below.

At present, the physical mechanism by which a DDT in degenerate supernova
material occurs is an area of current research (see  Refs.\ ~\citep{roepke07, 
Seitetal09, Woosetal09, schmidtetal10,PoludnenkoEtAl2011,cc2013} and references therein).
Simulations of supernovae involving DDT assume it occurs via
the Zeldovich-gradient mechanism~(Ref.\ \citealp{KhokOranWhee97}, but see also
 Refs.\ \citealp{Niem99,Seitetal09}),
in which a gradient in reactivity leads to a series of explosions that are in phase
with the velocity of a steadily propagating detonation wave.
Many authors suggest that when the flame reaches a state of distributed burning, which is
when turbulence on scales at or below the laminar flame width is fast enough to dominate
transport processes (see e.g., \citep{pope87}), fuel and ash are mixed and the temperature of
the fuel is raised and ``prepared'' in such a way to produce the required reactivity
gradient. A requirement for distributed burning is that the ratio of turbulent intensity to
the laminar flame speed must exceed some unknown threshold, which is still actively
researched~\citep{NiemWoos97, KhokOranWhee97,GoloNiem05,
roepkeandhillebrandt2005,Aspdetal08,Aspdetal10,poludnenkooran2010,poludnenkooran2011}.
Entrance into the distributed burning regime does not guarantee such a reactivity gradient
to form. \citet{Woosley07} and~\citet{Woosetal09} studied incorporating more stringent
requirements for these conditions to be met.

For burning under conditions found in the white dwarf, 
the ratio of turbulent intensity to laminar flame speed changes most rapidly due to
the change in laminar flame properties, which are strongly dependent on fuel density.
Therefore, a DDT is typically assumed to occur at a range of densities that 
generally lie in the range of $10^{6.7}$ to $10^{7.7}$~g~cm$^{-3}$~\citep{KhokOranWhee97,
LisewskiEtAl00, Woosley07, RoepNiem07, bravGarc08, maedaetal2010, jacketal2010},
although some research implements DDT criteria that include the strength of
background turbulence \citep[][and references therein]{seitenzahletal11}.
In simulations, the choice of  \rhoDDT\ increases or decreases the duration 
of the deflagration phase, which increases or decreases the amount of expansion 
prior to the detonation and hence the composition of the yield of Iron 
Group Elements (IGEs) \citep{jacketal2010}.

\section{Contemporary Research in the DDT Paradigm}
\label{sec:contemporary}

As outlined above, DDT models agree well with observations
and accordingly have been accepted as the favored model of
Type Ia supernovae. As the models work well, contemporary research
goes well beyond questions of whether or not explosions happen to
explore secondary parameters beyond the primary light curve parameter,
the brightness or $\Delta m_{15}(B)$ \citep{hoeetal2010,seitenzahletal11}.
In this section, we address the role of composition and central density
on the explosion of models of Type Ia supernovae in the single degenerate 
DDT paradigm.

\subsection{The Role of Progenitor Composition}
\label{sec:composition}
%

The composition of a white dwarf can influence the explosion in a host 
of ways, including the density at which the deflagration ignites, the 
flame speed, the density at which the DDT occurs, the energy released
during the phases of burning, and the neutron excess of the burned material,
which strongly influences properties of the light curve of an event, including
the peak brightness, expansion velocities, and the width-luminosity relation.
The most important
constituents of the progenitor white dwarf are \C{12}, \0{16}, and \Ne{22}.
The composition is principally set by the post-main sequence evolution 
\citep{DomiHoefStra01}, with the inner core formed during the convective 
core He-burning phase, and the outer layers formed in shell burning 
on the asymptotic giant branch \citep[and references therein]{Straetal03}. 
These two burning stages operate in very different burning regimes with core
He burning occurring at lower temperatures and higher densities than shell burning.
At these lower temperatures in the core, the \COreac\ reaction is favored over
$3\alpha$, and as a result \C{12} is depleted in the core, while it is not
in the outer layers~\citep{mengyang2011}. 

Additionally, the zero-age main sequence (ZAMS)
mass and metallicity influence the composition. As the ZAMS mass is increased
from the low end of the range that produce C-O white dwarfs, the central temperature
is increased during core He burning to favor $3\alpha$ over \COreac,
thereby increasing the C/O ratio. As the ZAMS mass continues to increase, however,
convection in the core becomes a dominant process
in transporting heat away from the core and the competition between $3\alpha$ and \COreac\
shifts back to favor \COreac, resulting in a reduced C/O ratio for the highest ZAMS masses
capable of producing CO white dwarfs. The C/O ratio is sensitive to the treatment of the
convection process and is still actively researched~\citep{DomiHoefStra01,Straetal03,
mengyang2011}. Increased metallicity increases the opacity and results
in a smaller core mass and lower central temperature, thus mimicking the effects of a
decreased ZAMS mass~\citep{umeda.nomoto.ea:evolution}. For a given temperature, increased metallicity
also reduces the $3\alpha$ rate to favor \C{12} destruction via \COreac, which results in
a lower C/O ratio; albeit, the effect is negligible for $Z < 0.02$~\citep{mengyang2011}.

Neutron rich isotopes are synthesized in two epochs. First, during He burning,
the aboriginal C, N, and O is transformed into \Ne{22}
\citep{timmes.brown.ea:variations}, 
leading to a direct dependence on metallicity of the parent stellar 
population. In addition, neutron-rich material is formed in the
pre-explosion convective C burning core at a comparable 
abundance~\citep{PiroBild08,Chametal08}.

The density at which the ignition of the deflagration occurs
is sensitive to several factors, including \C{12} abundance in the 
core and the thermal history of the core, which follows from the 
accretion rate and possibly properties of the He flashes \citep{Nomo84}.
Either of these may have metallicity dependencies involved with both the 
evolution of the parent stellar population and the still incompletely 
understood \citep{BranchLivioetal95} process of progenitor system formation.
There are also uncertainties in the screening enhancement of
nuclear reactions at these high densities
\citep{Gasques2005Nuclear-fusion-,YakoGasqetal06}, and in the reaction
rates themselves, particularly the $\mathrm{^{12}C + ^{12}C}$ reaction 
rate \citep[][and references therein]{bravoetal2011}.  

The energy per unit mass released during thermonuclear burning
influences both the dynamics of the white dwarf expansion 
during the deflagration phase and the dynamics of local buoyancy 
effects that accelerate the flame during the deflagration phase.
The composition influences the energy release of thermonuclear
burning in two ways. First, the gross nuclear energy available per 
unit mass is mostly sensitive to the abundance of \C{12}. 
Second, by changing the balance of protons and neutrons, the 
\Ne{22} abundance influences the ``ash'' to which the material burns,
with neutron rich material favoring more tightly bound nuclei and therefore
releasing more energy \citep{Caldetal07}. The abundances over a broad 
region of the white dwarf are involved in this case, and, accordingly,
involve the products of both the He core and shell burning.

The deflagration propagates via thermal conduction and the laminar flame
speed is sensitive to both the \C{12} and \Ne{22} abundances that 
influence the energy release and the speed of
the early stages of the nuclear fusion \citep{timmes_1992_aa,Chamulak2007The-Laminar-Fla}.  
Turbulence from the simmering phase and resulting from buoyancy instability 
as the flame propagates has the ability to add a nearly arbitrary amount of 
local area to the flame surface. The effect is to make the burning 
rate largely independent of the laminar propagation speed \citep{NiemHill95}.  
The early deflagration phase, however, when the laminar flame speed interacts 
with the lower strength turbulence from core convection sets the morphology 
of the burned region at the point when strong buoyancy effects take 
over \citep{ZingDurs07}. 

Although the physics of DDT are incompletely 
understood, the process will depend on composition because both the 
flame speed and width depend on the abundances of \C{12} and \Ne{22}. 
Depending on the details of the DDT, the transition could occur
either in material with the core composition or with the outer layer
composition due to the nature of subsonic burning and the response
of the star.
\figref{fig:ddt_4panel} shows four snapshots of a two-dimensional
simulation in which DDT is triggered at a specified density of 
$10^{7.1} \grampercc$ (green
contour) in a progenitor white dwarf with a different composition
in the core than the outer layer (marked by the blue contour). As
the deflagration ensues, the star expands and the density contour
shrinks. For this transition density, DDT occurs in material
from the core, but for lower transition densities, it may be possible
for DDT to occur in material from the outer layer.
The core \C{12} abundance will be lower than the outer layer due 
to the differences in temperature experienced in shell burning
as opposed to core He burning. The neutron excess in the core is
also expected to be greater than the outer layer due to the slow
C burning phase as well as potential sedimentation effects
\citep{bravoetal92,PiroBild08,Chametal08,althausetal2010,bravoetaldiff2011}.
Because the DDT density determines the duration of the deflagration
phase and thus degree of expansion of the star at the detonation, the influence 
of metallicity on the DDT density plays a significant role
in the outcome of the explosion \citep{Chametal08,jacketal2010}.

In addition to the indirect effect on energy release described above,
the \Ne{22} content, which sets the neutron excess, has a 
direct influence over the final nucleosynthetic products, particularly the
amount of \Ni{56}.  \citet{timmes.brown.ea:variations} showed that the
decrease in the \Ni{56} produced in the explosion, absent other factors,
should be linearly proportional to the \Ne{22} content and therefore the
original metallicity of the stellar population, a result seen in some
multidimensional results \citep{kasenroepkewoosley2009}.  The distribution of \Ne{22}
within the white dwarf is important, with \Ne{22} in the interior of the 
star influencing the gross yield and \Ne{22} in surface layers influencing
the ejecta opacity. 
Neutron excess also influences the structure of 
a white dwarf because it is supported by the pressure of degenerate electrons.  
The neutron excess, determined by the \Ne{22} abundance, sets the weight of baryons 
each electron must support. Accordingly, white dwarfs of a given mass with 
a higher neutron excess will be more compact because there are fewer total 
electrons.  Conversely, at the same central density, a star with more \Ne{22} 
and thus more neutrons, will be slightly less massive. These concurrent events on 
the density and structure of the white dwarf may be small \citep{HoefWheeThie98}, 
but should be included in realistic models due to the marginally bound nature of a 
near-Chandrasekhar mass white dwarf.

In a comprehensive study using one dimensional DDT models, \citet{hwt98} 
addressed the influence of initial composition on the nucleosynthesis,
light curves, and spectra along with cosmological implications. They found
a decreasing C/O ratio released less energy during burning, which increased
the duration of the deflagration phase, which in turn decreased the 
production of \Ni{56} and influenced both the bolometric and monochromatic 
light curves. They also varied the DDT transition density in one
model, and found that reduction of the transition density produced similar
results on the yield to decreasing the C/O ratio. They reported, 
however, that there are differences in the distribution of Si in the
expanding remnant that lead to slight differences in the light curves. 
\citet{hwt98} also reported that increasing
the metallicity of the progenitor decreased the yield of \Ni{56} because the
decreased number of protons led to the production of relatively more 
stable Fe-group nuclei. 

While one-dimensional simulations by \citet{umeda.nomoto.ea:origin} also found that a 
reduced C/O ratio decreased the production of \Ni{56}, later three-dimensional
deflagration-only
simulations by \citet{RoepHill04} found that the C/O ratio does not significantly
influence the final yield of \Ni{56}. They
attribute this to a complex interplay between the nucleosynthesis and nonlinear
turbulent flame evolution. This result was later confirmed with higher resolution
simulations and post-processing tracer particles with a more detailed nuclear reaction
network~\citep{Roepetal06_2}; however, due to the highly non-linear nature of the
turbulent flame propagation, more simulations with different ignition conditions
are necessary to rule out any systematic trend. The effect of the C/O ratio on
DDT has yet to be explored in detail; although, \citet{umeda.nomoto.ea:origin} speculated
that a larger C/O ratio should increase the transition density, and hence, increase
the yield of \Ni{56}. However, if DDT occurs due to the Zeldovich-gradient
mechanism in the distributed burning regime, then the transition density should
increase with decreasing C/O ratio because of the stronger effect on the
laminar flame properties~\citep{Chamulak2007The-Laminar-Fla,bravo10,jacketal2010}.

The results of \citet{hwt98} contain an important point concerning DDT models 
we wish to stress. The amount of expansion of the white dwarf during the deflagration 
phase significantly influences the explosion outcome, particularly the mass 
of Fe-group elements. We will quantify this result below in the discussion
of our research into the the role of the DDT transition density.

\citet{hoeetal2010} addressed secondary parameters of 
Type Ia supernova light curves, meaning parameters other than 
$\Delta m_{15}(B)$,
that influence the brightness of an event.
They argued that high quality light curves obtained by the Carnegie 
Supernova Project \citep{Contrerasetal2010,Folatellietal2010} show clear
evidence of the existence of secondary parameters responsible for variations
in the light curves and explored the relationship between possible parameters,
central density, the C/O ratio (directly following from the main
sequence mass), and metallicity, and the light curves from explosion models. 
The study was based
on DDT models with a fixed transition density of $2.3 \times 10^7 \grampercc$. 
They reported that central density and the C/O ratio may be treated as independent 
secondary parameters characterizing the light curve, and that larger main
sequence masses (lower  C/O ratios) produces a slower rise of the light curve.

\subsection{The Role of Central Density}
\label{sec:central}

As described above, the composition of the progenitor white dwarf
can play an important role in the outcome of an explosion. The composition
is determined by many factors, such as the mass of the main sequence star
that became the white dwarf, but it is also influenced by properties
of the host galaxy such as metallicity.  These properties
lead to correlations between galaxy type and brightness of
events. A similar property of the progenitor white dwarf that influences 
the brightness and follows from the age of the progenitor system,
and hence the age of the galaxy, is the central density.

As the accreting white dwarf approaches the Chandrasekhar mass, it is
relatively insensitive to variations in the central density because of the
degenerate equation of state. The configuration of the convecting interior is
uncertain, however, because of an incomplete understanding of the physics that
influence ignition. There have been many studies of the convective phase 
\citep{Hoflich2002On-the-Thermonu, WoosWunsKuhl04, Kuhletal06, PiroChan08,
PiroBild08, zingaleetal2009, zingaleetal2011, nonakaetal2012}, but examples of
remaining uncertainties include the \C{12}--\C{12} rate and its influence on
the ignition density \citep{JianRehmetal07, CoopSteiBrow09,
iapichinolesaffre2010, bravoetal2011} and the loss of energy due to neutrinos,
i.e.\ the convective Urca process \citep{paczynski72, Iben78a, Iben78b, Iben82,
Stein1999The-Role-of-Kin, Lesaffre2005A-two-stream-fo}. 

While these many questions remain, it is accepted that the initial mass of
the white dwarf (i.e., the mass at which it formed) and the accretion history,
particularly the period of cooling prior to the onset of accretion, largely
determine the state of the interior of the white dwarf~\citep{seitenzahletal11}.
The study of \citet{Lesaffre2006The-C-flash-and} found that white dwarfs experiencing
a long cooling period prior to the onset of accretion cool to a lower
central temperature, so that once accretion begins to reheat the core a greater
increase in density is necessary to reach the conditions for a thermonuclear
runaway (see especially Figure~3 of Ref.\ \citealp{Lesaffre2006The-C-flash-and}).
Therefore, while the central density of the white dwarf may have been the same
prior to the cooling period, at ignition of the thermonuclear runaway the
central density is higher for white dwarfs with a longer cooling time.

The effect of density on electron capture rates has been known
for some time \citep[][and references therein]{arnett69}. Electron capture
rates increase with density, so that at the higher densities found in
near-Chandrasekhar mass white dwarfs, the rates are rapid enough that
significant neutronization occurs during the explosion. 
The issue of the effect of central density has been explored by many
research groups~\citep{Bravo1990, Roepetal06_2, fisheretal10,
hoeetal2010,seitenzahletal11} and
below we highlight some of this work. 

\citet{hoeetal2010} in their study of secondary
parameters of Type Ia supernova light curves considered central density.
They found that the central density influences the later evolution of
the light curve ($\sim 30$ days after maximum and later) with increases
in central density causing the later portion of the light curve to shift
down and vice versa. They attribute this difference to differences in
the amount and distribution of \Ni{56} in the central envelope.

Recently \citet{seitenzahletal11} explored the effect of changing central
density on
the explosion outcome with three-dimensional models and a treatment of the
energetics of the thermonuclear burning similar to the method we employ
(described
below) that includes weak interactions, specifically electron capture, allowing
their models to neutronize. \citet{seitenzahletal11} found that the increased
neutronization rates decreased the relative abundance of \Ni{56} in the Fe-group 
material synthesized in the explosion. Considering an identical spatial
distribution
of ignition kernels, however, they found the increased gravitational
acceleration following
from a higher central density increases the production of turbulence, which
produces
a higher burning rate and triggers the DDT sooner, with less expansion of the
star and therefore a higher yield of Fe-group elements. These two effects
combine
to give the relatively constant yield of \Ni{56}, and they conclude that
the net effect is that the mass of \Ni{56} synthesized in an explosion appears
insensitive to the central density of the progenitor white dwarf. More recent
three-dimensional results including detailed nucleosynthetic post-processing of $1 \times 10^6$ 
tracer particles per model suggest that \Ni{56} mass increases with central density. 
In those models, the fraction of \Ni{56} of IGEs decreases due to neutronization, 
but the increase in the total amount in IGE outweighs the decrease due to 
neutronization \citep{seitenzahletal12}.

\citet{seitenzahletal11} also report that the \Ni{56} yield is sensitive to the
initial
conditions, which largely determine the behavior of the deflagration phase. They
found that a strong deflagration phase following from a many-kernel (larger)
ignition region
produces a lower yield of \Ni{56} primarily because of the increased expansion
of the 
star during the deflagration phase but also due to increased neutronization
during 
the deflagration phase. They note that this confirms previous studies indicating
that the strength of the deflagration phase is a primary parameter of \Ni{56}
production \citep{roepkeniemeyer2007}.

\section{Research Results}
\label{sec:research}

In this section we provide a brief summary of our own 
theoretical research into systematic effects on the brightness
of Type Ia supernova explosions. We describe our methodology,
including our models and statistical approach to studying systematic
effects, and present highlights from our investigation into the 
role of the metallicity and central density of the progenitor white dwarf.
We gratefully acknowledge contributions to the results presented here
from our many collaborators.

\subsection{Methodology}

Our methodology for the theoretical study of Type Ia 
supernovae consists of four principal parts. First is
the ability to construct parameterized, hydrostatic initial white
dwarf progenitors that can freely change thermal and compositional
structure to match features from the literature about
progenitor models~\citep{jacketal2010}. Second is a model flame and energetics
scheme with which to track both (subsonic) deflagrations and (supersonic)
detonations as well as the evolution of dynamic ash in nuclear statistical
equilibrium (NSE).
This flame/energetics scheme is implemented
in the Flash hydrodynamics
code~\citep{Fryxetal00,calder.curtis.ea:high-performance,calder.fryxell.ea:on}.
Third is utilization of a scheme to post-process the density and temperature
histories of Lagrangian tracer particles with a detailed nuclear
network in order to calculate detailed nucleosynthetic
yields~\citep{brown.calder.ea:type,townetal10}. Fourth, we developed a
statistical framework with which to perform ensembles of simulations for
well-controlled studies of systematic effects.
Below we highlight the flame model and the statistical
framework. Complete details of the methodology can be found in
previously published results~\citep{brown.calder.ea:type,Caldetal07,townsley.calder.ea:flame,townetal2009,townetal10}.
The simulations presented here are all two-dimensional.

\subsubsection{Flame Model}

The significant disparity between the length scale of a white
dwarf ($\sim 10^9\ensuremath{\;}{\ensuremath{\mathrm{cm}}}$) and the width
of laminar nuclear flame ($< 1\ensuremath{\;}{\ensuremath{\mathrm{cm}}}$)
necessitates the use of a model flame in simulations of Type Ia 
supernovae. Even simulations with adaptive mesh refinement cannot
resolve the actual diffusive flame front in a simulation of the event.
The model we use propagates an artificially
broadened flame front with an advection-diffusion-reaction (ADR)
scheme~\citep{Khok95,VladWeirRyzh06} that has been demonstrated to be
acoustically quiet and produce a unique flame speed~\citep{townsley.calder.ea:flame}.
This scheme evolves
a reaction progress variable $\phi$, where $\phi=0$ indicates unburned fuel
and $\phi=1$ indicates burned ash, with the
advection-diffusion-reaction equation
\begin{equation}
  \label{eq:ard}
  \partial_t \pv + \vec{u}\cdot\nabla \pv = \kappa \nabla^2 \pv + 
\frac{1}{\tau} R\left(\phi\right).
\end{equation}
Here $\kappa$ is a constant and $R(\phi)$ a non-dimensional function, and both
are tuned to propagate the reaction front at the physical speed of the
real flame~\citep{timmes92,Chamulak2007The-Laminar-Fla} and to be just wide
enough to be resolved in our simulation.  We use a modified version of the
KPP reaction rate discussed by~\citet{VladWeirRyzh06}, in which
$R\propto(\phi-\epsilon)(1-\phi+\epsilon)$, where $\epsilon \simeq 10^{-3}$,
which is acoustically quiet and gives a unique flame speed
\citep{townsley.calder.ea:flame}. 

In simulating the deflagration phase, the flame front separates expanded
burned material (the hot ash) from denser unburned stellar material (cold
fuel). The expansion and buoyancy of the burned material forces the interface
upward into the denser fuel, and the configuration is susceptible to
the Rayleigh-Taylor fluid instability~\citep{taylor+50,chandra+81}. It is necessary to enhance the
flame speed in order to prevent turbulence generated by the Raleigh-Taylor
instability from destroying the artificially broadened flame front.
In the simulations discussed here, the enhancement is accomplished by the method
suggested by \citet{Khok95} in which we prevent the flame front speed, $s$,
from falling below a threshold that is scaled with the strength of
the Raleigh-Taylor instability on the scale of
the flame front ($s\propto \sqrt{g\ell}$, where $g$ is the local gravity and
and $\ell$ is the width of the artificial flame, which is a few times the
grid resolution).  This scaling of the flame speed prevents the Rayleigh-Taylor
instability from effectively
pulling the flame apart and also mimics what is a real enhancement of the burning
area that is occurring due to structure in the flame surface on unresolved scales.
It is expected that, for much of the white dwarf deflagration, the flame is
``self-regulated'', in which the small scale structure of the flame surface
is always sufficient to keep up with the large-scale buoyancy-driven motions
of the burned material.  Thus the actual burning rate is determined by this
action, which is resolved in our simulations \citep{townsley2009}.

This technique has demonstrated a suitable level of convergence for studies of 
Type Ia supernovae \citep{gamezo.khokhlov.ea:thermonuclear,townsley.calder.ea:flame}.
The technique explicitly drops the flame speed to zero below a density of
$10^7$ g cm$^{-3}$, approximately where the real flame will be extinguished.
This prescription captures
some effects of the Rayleigh-Taylor instability and maintains the integrity
of the thickened flame, but it does not completely describe the flame-turbulence
interaction. Also, we neglect any enhancement from background turbulence from
convection prior to the birth of the flame. Future work will include
physically-motivated models for these effects~\citep{jackinprep}.

In addition to the model flame, the energetics scheme also includes the nuclear 
energy release occurring at the flame front, in subsequent burning stages, and
in the dynamic ash in nuclear statistical equilibrium. We performed a detailed study of 
the nuclear processes occurring in a flame in the interior of a white dwarf and developed 
an efficient and accurate method for incorporating the results into numerical 
simulations~\citep{Caldetal07,townsley.calder.ea:flame,townetal2009}.
Tracking even tens of nuclear species is computationally prohibitive, and many
more than this are required to accurately calculate the physics such as electron
capture rates that are essential to studying rates of neutronization.  We instead
reproduce the energy release of the nuclear reactions with a highly abstracted model
based on tabulation of properties of the burned material calculated in
our study of the relevant nuclear processes.
This method accurately captures the thermal history of the material
as it burns and evolves, which enables embedded particles to obtain
accurate Lagrangian density and temperature histories. Detailed
abundances can then be recovered by post-processing these time
histories with a nuclear network including hundreds of nuclides.

The nuclear processing can be well-approximated as a three stage process:
initially C is consumed, followed by O, which creates a
mixture of Si group and light elements that is in
quasi-statistical equilibrium, also known as nuclear statistical 
quasi-equilibrium 
(NSQE)~\citep{bodclafow1968,woosleyarnettclayton73,ifk1981,khok1981,khok1983}); finally
the Si-group nuclei are converted to
Fe group, reaching full nuclear statistical equilibrium, NSE.
In both of these equilibrium states, the capture and creation of light elements
(via photodisintegration) is balanced, so that energy release can continue by
changing the relative abundance of light (low nuclear binding energy) and heavy
(high nuclear binding energy) nuclides, an action that releases energy as buoyant burned
material rises and expands.
We track each of these stages with separate progress variables and separate
relaxation times derived from full nuclear network calculations
\citep{Caldetal07}.  We define three progress variables
representing consumption of C, $\phi_{\rm C}$, consumption of O to
material in NSQE, $\phi_{\rm NSQE}$, and conversion of Si to Fe-group nuclides
to form true NSE material, $\phi_{\rm NSE}$.  The physical state of the fluid is
tracked with the electron number per baryon, $Y_e$, the number of nuclei
per baryon, $Y_{\rm i}$, and the average binding energy per baryon, $\bar q$,
the minimum properties necessary to hydrodynamically evolve the fluid.
C consumption is coupled directly to the flame progress variable,
$\phi_{\rm C}\equiv\phi$, from eq.\ (\ref{eq:ard}) above, and the later flame
stages follow using simple relationships from more detailed calculations.


Burning and evolution of post-flame material change the nuclear binding
energy, and we use the binding energy of magnesium to approximate
the intermediate burning products of C. The method is finite
differenced in such a way to ensure explicit conservation of energy.
Weak processes that neutronize the material (e.g.\  electron capture) are 
included in the calculation
of the energy input rate, as are neutrino losses, which are calculated
by convolving the NSE distribution with the weak interaction cross sections.
Tracking the conversion of Si-group nuclides to the Fe-group is important
for studying the effects of electron capture because the thresholds are lower for
the Fe-group nuclides.  Both the NSE state and the electron capture rates were
calculated with a set of 443 nuclides including all which have weak interaction
cross sections given by \citet{langanke.martinez-pinedo:weak}.  

Electron capture feeds back on the hydrodynamics in
three ways: the NSE can shift to more tightly bound elements as the electron
fraction, $Y_e$, changes, releasing some energy and changing the local
temperature; also the reduction in $Y_e$ lowers the Fermi energy, reducing
the primary pressure support of this highly degenerate material and having an
impact on the buoyancy of the neutronized material; finally neutrinos are
emitted (since the star is transparent to them) so that some energy is lost
from the system. Also, as we will see below, increased rates of neutronization
produce yields of more neutron rich material and therefore less \Ni{56}, thereby
influencing the brightness of an event. Complete
details of the NSE calculations may be found in Ref.\ \citet{SeitTownetal09} and
the details of the implementation in our simulations may be found in
\citet{Caldetal07,townsley.calder.ea:flame,townetal2009}.

In addition to all of these effects during the deflagration phase
of Type Ia supernovae, the progress-variable-based method has been extended to
model the gross features of detonations~\citep{Meaketal09,townetal10}.
Instead of coupling the first burning stage, $\phi_{\rm C}$,
representing C consumption, to the ADR flame
front, we instead can use the actual temperature-dependent rate for C
burning, or possibly a more appropriate effective rate.  Doing so allows  shock
propagation to trigger burning and therefore create a propagating detonation.
This method has been used successfully by Ref.\ \citep{GameKhokOran05}
in modern studies of the DDT, and our
multistage burning model shares many features with theirs (see also
Refs.\ \citep{khokhlov91+dd} and \citep{khokhlov+00}).  We treat the later
burning stages very similarly, though we have taken slightly more care to
track the intermediate stages and have nearly eliminated acoustic noise when
coupling energy release to the flame.

\subsubsection{Statistical Framework}

We also developed a theoretical framework for the study of systematic effects
in Type Ia supernovae \citep{townetal2009}.
This framework allows the evaluation of the average dependence of the properties
of supernovae on underlying parameters,
such as composition, by constructing a theoretical sample based on a
probabilistic initial ignition condition.  Such sample-averaged dependencies are
important for understanding how Type Ia models may explain features of the
observed sample, particularly samples generated by large dark energy
surveys utilizing Type Ia supernovae as distance indicators.

The theoretical sample is constructed to represent statistical properties of
the observed sample of Type Ia supernovae such as the mean inferred $\Ni{56}$
yield and variance. Within the DDT paradigm, the variance in $\Ni{56}$
yields can be explained by the development of fluid instabilities during
the deflagration phase of the explosion. By choice of the initial
configuration of the flame, we may influence the growth of these fluid
instabilities resulting in varying amounts of $\Ni{56}$ synthesized
during the explosion. In \citet{townetal2009}, we found that perturbing a
spherical flame surface with radius ($r_0 = 150$ km) with spherical
harmonic modes ($Y_l^m$) between $12 \leq l \leq 16$ with random 
amplitudes ($A$) normally distributed between $0-15$ km and, for 
three-dimensional simulations, random phases ($\delta$)
uniformly distributed between $\mbox{-}\pi-\pi$ best characterized the
mean inferred $\Ni{56}$ yield and sample variance from observations:
\begin{eqnarray}
\label{eq:init_surf}
r(\theta) &=& r_0 + \sum_{l=l_{\rm min}}^{l_{\rm max}} A_l Y_l(\theta) \\
r(\theta,\phi) &=& r_0 + \sum_{l=l_{\rm min}}^{l_{\rm max}} \sum_{m=-l}^l 
\frac{A_l^m e^{i\delta_l^m}}{\sqrt{2l+1}} Y_l^m(\theta,\phi)
{\rm .}
\end{eqnarray}
With a suitable random-number generator, a
sample population of progenitor white dwarfs is constructed by defining the initial
flame surface for a particular progenitor.

\subsection{Investigation into Metallicity}

We investigated the
direct effect of reprocessed stellar material (metals) in the host
galaxy via the initial neutron excess of the progenitor white dwarf
in~\citet{townetal2009}.
Because of weak interactions, metals produced by nuclear burning are
more neutron rich than H and He, and, accordingly, the
neutron excess of these elements is thought to drive the explosion
yield toward stable Fe-group elements. Thus, there is relatively less
radioactive $^{56}$Ni in the NSE mix, which results in a dimmer
event~\citep{timmes.brown.ea:variations}.
We investigated this effect by introducing $^{22}$Ne into the
progenitor white dwarf as a proxy for (neutron-rich) metals.
The presence of $^{22}$Ne influences the progenitor structure, the energy
release of the burn, and the flame speed.  The study was designed to measure
how the $^{22}$Ne content influences the competition between rising plumes
and the expansion of the star, which determines the yield.
We performed a suite of 20 DDT simulations varying only the
initial $^{22}$Ne in a progenitor model, and found a
negligible effect on the pre-detonation expansion of the star
and thus the yield of NSE material.  The neutron excess sets the
amount of material in NSE that favors stable Fe-group elements over
radioactive $^{56}$Ni. Our results were consistent with earlier
work calculating the direct modification of $^{56}$Ni mass from
initial neutron excess~\citep{timmes.brown.ea:variations}.

In \citet{jacketal2010} we expanded the
earlier study~\citep{townetal2009} to include the indirect 
effect of metallicity in
the form of the $^{22}$Ne mass fraction through its influence
on the density at which the DDT takes place. We simulated 30 realizations
each at 5 transition densities between $1-3\times10^7$~g~cm$^{-3}$ for a
total of 150 simulations.  We found a quadratic dependence of the NSE
yield on the log of the transition density, which is determined by the
competition between rising unstable plumes and stellar expansion. By
then considering the effect of metallicity on the transition density, we
found the NSE yield decreases slightly with metallicity, but that the ratio
of the $^{56}$Ni yield to the overall NSE yield does not change as
significantly. Observations testing the dependence of the yield on metallicity
remain somewhat ambiguous, but the dependence we found is comparable to that
inferred from~\cite{bravo10}.  We also found that the scatter in the results
increases with decreasing transition density, and we attribute this increase in
scatter to the nonlinear behavior of the unstable rising plumes.

Figure~\ref{fig:aaron} illustrates these results by plotting mass of \Ni{56} 
vs.\ metallicity for our models and the observational results of \citet{konishietal2011}. 
The constant DDT density results (blue curve) represent the direct effect of introducing 
$^{22}$Ne into the progenitor white dwarf as a proxy for (neutron-rich) metals. 
The red curves present results adapted from \citet{jacketal2010} that include
the indirect effect of metallicity influencing DDT density. The results inferred from
\citet{konishietal2011} are the black points. In the plot, metallicity is quantified by
comparing the ratios of Fe to H in the progenitor white dwarfs relative to
our Sun~\citep{co2007}. We write this relationship as 
\begin{displaymath} 
\left[{\rm Fe}/{\rm H} \right] 
\equiv \log_{10} \left[ \frac{\left(N_{\rm Fe}/N_{\rm H}\right)_{\rm star}}
{\left(N_{\rm Fe}/N_{\rm H}\right)_{\rm Sun}} \right] \;.
\end{displaymath}

\subsection{Investigation into Central Density}

We investigated the effect of central density on the explosion yield
in  \citet{Krueger2010On-Variations-o}. While we found little change 
in the overall production of NSE material, we found a definite trend 
of decreasing $^{56}$Ni production with increasing progenitor central 
density. We attribute this result to higher rates of weak interactions
(electron captures) that produce a higher proportion
of neutronized material at higher density.  More neutronization
means less symmetric nuclei like $^{56}$Ni, and, accordingly,
a dimmer event. This result may explain
the observed decrease in Type Ia supernova brightness with increasing delay
time. The central density of the progenitor is determined by
its evolution, including the transfer of mass from the companion.
If there is a long period of cooling before the onset of mass transfer,
the central density of the progenitor will be higher when the core
reaches the C ignition temperature~\citep{Lesaffre2006The-C-flash-and},
thereby producing less $^{56}$Ni and thus a dimmer event.  In addition, we found
considerable variation in the trends from some realizations, stressing the
importance of statistical studies.

In \citet{kruegetal12} we expanded our earlier investigation into the role of
central density on the production of \Ni{56} and hence the brightness of an
event. We presented the details of our models, and found that the distribution 
of \Ni{56} in our models typically shows a somewhat clumpy concentration in
the inner region of the remnant, with some dependence of the morphology on the
central density. We found a trend of a shortened period of deflagration at higher
central densities, but owing to increased rates of neutronization, the total
neutronization is greater at higher densities, which explains our 
trend of decreasing $^{56}$Ni production with increasing progenitor central 
density. We also considered main sequence evolution in comparing the masses
of \Ni{56} produced to observations thereby obtaining an improved brightness-age 
relation from our results.

Figure \ref{fig:neill} presents results adapted from  \citet{kruegetal12}
relating the results to observations. The principal result
of a simulation is the mass of \Ni{56} produced, which is directly related to 
the brightness of an event. Using the method outlined in \citet{howelletal+09}, we 
can convert \Ni{56} masses to stretch 
values, a measure related to peak brightness~\citep{Goldhaber2001Timescale-Stret},
for a more direct comparison with observations.  We combined the 
central density values with the results of \citet{Lesaffre2006The-C-flash-and}, which
correlate the central density at the time of the ignition of the flame front to
the cooling time of the progenitor WD (that is, the period of isolation prior to
the onset of accretion), allowing us to express our results as
ages.  The results of \citet{Lesaffre2006The-C-flash-and} suggest that a WD with a central
density of $1\times10^9$~g~cm$^{-3}$ will not ignite without further accretion,
so for this comparison simulations with that value of central density were neglected.
We also applied a shift to our data to account for the main sequence age, $\tau_{\rm MS}$, which we
take to be constant across our results. We consider two estimated limiting values 
for $\tau_{\rm MS}$ (0.05 and 1.0~Gyr,
corresponding to main-sequence masses of approximately 8.0 and 1.5 solar masses,
respectively; see \citealt{HansenEtAl04}) and included them with the original
$\tau_{\rm MS} = 0$ result in \figref{fig:neill}.  Adding in a $\tau_{\rm MS}$
consistent with our C-O progenitors brings our results into better agreement
with the two right-most points of \citet{neilletal+09}.

Figure \ref{fig:neill} shows that our theoretical results are not in complete
agreement with the observed data.  Observationally, the age-brightness
correlation may flatten at young ages, while our data do not, resulting in
our data being too bright relative to young Type Ia supernovae. This study
isolated the effects
of central density and related that to age assuming a constant main-sequence
mass, but there are other effects that may be correlated.  Examples of such
potentially correlated effects are: main-sequence mass and its correlation
with central density, metallicity of the progenitor, core $^{12}$C fraction
prior to ignition of the deflagration, sedimentation, and others.  Inclusion of
such effects may modify the results presented here and are the subjects of
future work.

\section{Summary and Conclusions}
\label{sec:conclusions}

We have presented an overview of Type Ia supernova explosion models
currently under study and a brief description of contemporary research.
Both observations and theoretical models are at a level of sophistication 
that allows comparison at a far more detailed level than just bulk properties 
such as energy release or bolometric light curve. Instead, studies explore
secondary parameters of light curves and the role of properties such as 
the age and  composition of the progenitor system on the explosion yield and 
light curve. The structure and composition of a white dwarf follow
from properties of the host galaxy such as metallicity and age, and both
theorists and observers are able to study trends of brightness with
host galaxy type. These studies will better inform future observational
campaigns probing the expansion history of the Universe and the properties
of dark energy.

Certainly, however, the problem is far from solved. The problems of 
discerning the progenitor systems(s) and explaining the diversity of events 
remain. Researchers are actively exploring multiple progenitor systems, to 
explain ``normal'' events as well as outliers, both dim and bright. Modeling 
explosions in these progenitor systems is a challenging problem, necessitating
the development of complex sub-grid-scale models, and fundamental
questions remain about models and the basic physics therein, e.g.\ 
the unsolved problem of turbulent combustion. Good progress is being made, 
however, and our assessment is that many of the outstanding 
questions will be answered in the near future. 

We presented results from our research into systematic effects on the
production of \Ni{56} and hence the brightness of an event. Our simulations
assume the single degenerate progenitor system and that the explosion 
proceeds via  a DDT. Our approach to studying systematic effects is 
via suites of simulations allowing quantification of statistical properties. 
One important result from our research we stress here is the need for a 
statistical approach to discerning trends as illustrated by the relatively 
large standard deviation in Figure \ref{fig:aaron}. 

Our models that include the direct effect of metallicity indicate a weak
trend of decreasing brightness with increasing metallicity that is
consistent with earlier results~\citep{timmes.brown.ea:variations}.
Models including the indirect effect of metallicity on flame speeds
and the deflagration to detonation transition density demonstrate a
significantly stronger trend of decreasing brightness with increasing
metallicity. Our results indicate a quadratic dependence of the yield 
of \Ni{56} on the log of the transition density. Relating metallicity to 
the transition density gave relationship between brightness and 
metallicity~\citep{jacketal2010}.

We also presented results from our research into the effect of central
density. Our suites of simulations showed a strong trend of decreasing 
brightness with increasing central density that we attribute to increase 
rates of neutronization. Relating our central density results to the age
of the progenitor system allows comparison to observations, and in so
doing we were able to provide a theoretical explanation of the observed 
trend that Type Ia supernovae from older host galaxies are systematically 
dimmer~\citep{kruegetal12}. The agreement between our trend and the
observational trends is not perfect, however, and we recognize that
uncertainties abound and our two-dimensional models are incomplete.
Plans for the future include targeted three-dimensional simulations 
including more complete physics to confirm our results.  

Our conclusion is that this is an exciting time to be studying 
the complex problem of thermonuclear supernovae. Substantial progress 
is being made by both observers and theorists, and we hope that this
review of one part of the problem, the role of chemical composition
on explosion models, conveys some of this excitement. 
In preparing this review, we were struck by the remarkable progress 
that has been made in modeling this complex, multi-scale, multi-physics
problem by many talented scientists. We also hope this review conveys our
enthusiasm and respect for all of the careful study, both past and present, 
of the problem of Type Ia supernovae.

\acknowledgements

This work was supported by the Department of Energy through grants
DE-FG02-07ER41516, DE-FG02-08ER41570, and DE-FG02-08ER41565, and by NASA
through grant NNX09AD19G.  ACC also acknowledges support from the Department of
Energy under grant DE-FG02-87ER40317.  DMT received support from the Bart
J.\ Bok fellowship at the University of Arizona. This research
was performed while APJ held a National Research Council Research Associateship
Award at the Naval Research Laboratory. The authors also acknowledge the
hospitality of the KITP, which is supported by NSF grant PHY05-51164, during
the programs ``Accretion and Explosion: the Astrophysics of Degenerate Stars''
and ``Stellar Death and Supernovae.''  The software used in this work was in
part developed by the DOE-supported ASC/Alliances Center for Astrophysical
Thermonuclear Flashes at the University of Chicago.  This research utilized
resources at the New York Center for Computational Sciences at Stony Brook
University/Brookhaven National Laboratory which is supported by the
U.S.\ Department of Energy under Contract No.\ DE-AC02-98CH10886 and by the
State of New York. The authors gratefully thank Ivo Seitenzahl, Max Katz,
and Michal Simon for helpful comments on drafts of this
manuscript. The authors also thank the anonymous referee for
insightful comments and suggestions that significantly improved the manuscript.


%

\clearpage


\begin{figure}[tbh]
\includegraphics[angle=0,width=\columnwidth]{f01.eps}
\caption{\label{fig:ddt_4panel} A two-dimensional delayed-detonation simulation
in which DDT is triggered at a specified density of $10^{7.1} \grampercc$ shown
by the green contour. The core of the progenitor white dwarf is more carbon-depleted
than the outer layer with the core-outer layer boundary highlighted with the blue
contour. Four different snapshots in time are provided that show the evolution of
the flame and the expansion of the white dwarf. Note that the rightmost frame is
scaled by a factor of two relative to the others.
}
\end{figure}

\begin{figure}[tbh]
\includegraphics[angle=0,width=\columnwidth]{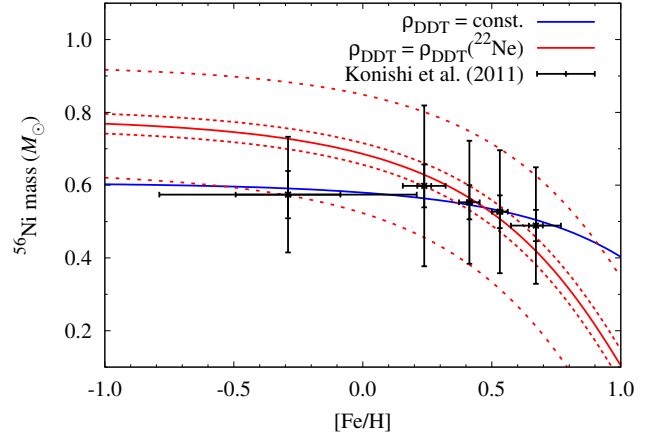}
\caption{\label{fig:aaron} Plot of  mass of $^{56}$Ni vs.\ metallicity comparing
simulations that varied DDT densities with metallicity 
to the observational results of \citet{konishietal2011} (black points). The constant DDT 
result, blue curve, assumed that the DDT density did not vary with metallicity but included other
metallicity effects on the structure of the progenitor star. The red curves present
results from adapted from \citet{jacketal2010}, with the 
the solid curve presenting the average
value of simulations at a given metallicity and the widely-spaced dashed and dashed curves
representing standard deviation and standard error, respectively.
}
\end{figure}

\begin{figure}[tbh]
   \includegraphics[angle=270,
      width=\columnwidth]{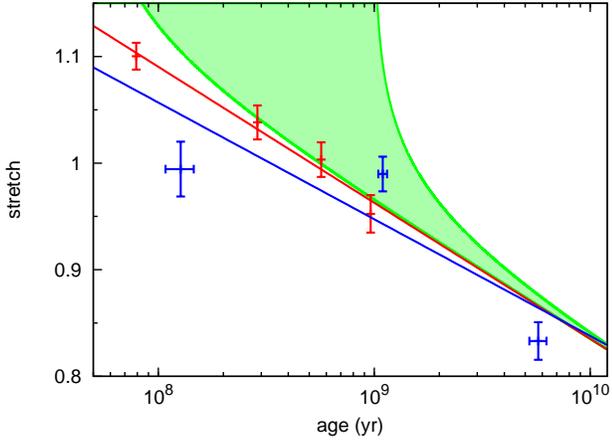}
   \caption{Plot of stretch vs.\ age comparing the scaled results of simulations
      varying the central density of the progenitor white dwarf 
      to the observations of \citet{neilletal+09}. In red are the points from this
      study with no shift (i.e., $\tau_{\rm MS} = 0$~Gyr), along with the
      standard error of the mean and a best-fit trend line. The green shaded 
      region shows our best-fit line with $\tau_{\rm MS} = 0.05 - 1.0$~Gyr.  
      In blue are the binned and averaged points from Figure~5 of \citet{neilletal+09}, 
      along with their best-fit trend line. Adapted from \citet{kruegetal12}.}
   \label{fig:neill}
\end{figure}

\end{document}